\documentclass[prd,aps,floats,floatfix,nofootinbib]{revtex4}
\usepackage{amsmath}
\usepackage{amsfonts}
\usepackage{graphicx}
\usepackage{slashed}
\usepackage{dcolumn}
\usepackage{bm}
\usepackage[dvips]{color}
\newcommand{\beq}{\begin{equation}}
\newcommand{\eeq}{\end{equation}}
\begin{document}
%%%%%%%%%%%%%%%%%
\newcommand{\vect}[1]{\overrightarrow{#1}}
\newcommand{\smbox}[1]{\mbox{\scriptsize #1}}
\newcommand{\tanbox}[1]{\mbox{\tiny #1}}
\newcommand{\vev}[1]{\langle #1 \rangle}
\newcommand{\Tr}[1]{\mbox{Tr}\left[#1\right]}
\newcommand{\picwidth}{3.4in}
\newcommand{\lae}{$\stackrel{<}{\sim}$}
\newcommand{\gae}{$\stackrel{>}{\sim}$}
\newcommand{\laem}{\stackrel{<}{\sim}}
\newcommand{\gaem}{\stackrel{>}{\sim}}
%%%%%%%%%%%%%%%%

%\preprint{MSU-HEP-080508}
\title{Axigluons cannot explain the observed top quark forward-backward asymmetry}

\author{R. Sekhar Chivukula}
\email[]{sekhar@msu.edu}
\author{Elizabeth H. Simmons}
\email[]{esimmons@msu.edu}
\author{C.-P. Yuan}
\email[]{yuan@pa.msu.edu}
\affiliation{Department of Physics,
Michigan State University, East Lansing, MI 48824, USA}
\date{\today}

\begin{abstract}
We study an $SU(3)^2$ axigluon model introduced by Frampton, Shu, and Wang to explain the recent Fermilab Tevatron observation of a significant positive enhancement in the top quark forward-backward asymmetry relative to standard model predictions.  First, we demonstrate that data on neutral $B_d$-meson mixing excludes  the region of model parameter space where the top asymmetry is predicted to be the largest.  Keeping the gauge couplings below the critical value that would lead to fermion condensation imposes further limits at large axigluon mass, while precision electroweak constraints on the model are relatively mild.  Furthermore, by considering an extension to an $SU(3)^3$ color group, we demonstrate that embedding the model in an extra-dimensional framework can only dilute the axigluon effect on the forward-backward asymmetry. We conclude that axigluon models are unlikely to be the source of the observed top quark asymmetry.
\end{abstract}

\maketitle

\section{Introduction}

Recent measurements of the forward-backward asymmetry ($A^t_{FB}$) in top quark pair production at the Fermilab Tevatron have shown a significant positive deviation from the small value predicted in the standard model.  The value reported by CDF, based on $3.2 fb^{-1}$ of integrated luminosity, is \cite{CDFvalue}
\begin{equation}
A^t_{FB} = 0.193 \pm 0.065 {\rm(stat)} \pm 0.024 {\rm (syst)}\,.
\end{equation}
This is consistent with previous measurements from D0 \cite{:2007qb}  and CDF \cite{Aaltonen:2008hc}, and is noticeably (about $2\sigma$) larger than the value of $A^t_{FB} = 0.051$ given by NLO QCD calculations   \cite{Antunano:2007da,Bowen:2005ap,Kuhn:1998kw,Kuhn:1998jr,Almeida:2008ug}.  A subsequent measurement by CDF with 5.3 $fb^{-1}$ continues to report an approximately $2\sigma$ deviation from QCD at NLO \cite{ICHEP2010}.

Numerous models have been proposed to explain this discrepancy \cite{Sehgal:1987wi,Choudhury:2007ux,Djouadi:2009nb,Martynov:2009en,Jung:2009jz,Cheung:2009ch,Shu:2009xf,Arhrib:2009hu,Ferrario:2009ee,Dorsner:2009mq,Jung:2009pi,Cao:2009uz,Barger:2010mw,Cao:2010zb,Xiao:2010hm,Martynov:2010ed,Ferrario:2010hm,Rodrigo:2010gm}.   Among them is an intriguing axigluon \cite{Frampton:1987dn} model, based on an $SU(3) \times SU(3)$ gauge group, proposed by Frampton, Shu, and Wang \cite{Frampton:2009rk}.   They write down a complete low-energy form of the model and suggest that it could conceivably arise from an extra-dimensional theory at higher energies.  They calculate the size of the effect on $A^t_{FB}$ and find a region of model parameter space where the model predicts an asymmetry value that is enhanced with respect to the QCD value, though its best level of agreement with the CDF measurement is at the $1\sigma$ level.  They also note that an axigluon producing such an effect would be light enough that the value of the asymmetry as a function of invariant top-pair mass would show a characteristic rise below and fall above the axigluon mass.

We explore the phenomenology of this model and find some significant constraints on the allowed values of the axigluon mass $M_C$ and the mixing angle, $\theta$, between the two strong gauge couplings.    First, we find that bounds imposed by data on neutral $B_d$-meson mixing exclude  the region of parameter space where the top asymmetry showed the greatest agreement with the data.\footnote{The recent analysis of \cite{Chen:2010wv} on  flavor-physics issues in this axigluon model  is rendered invalid by their inconsistent assumptions about the values of the vector and axial quark couplings.  Specifically, they set $g_A = -1.155 g_s$ and simultaneously set $g_V = g_A$; as discussed in \cite{Frampton:2009rk} and our Eq. (\ref{eq:couplingss}), the first choice corresponds to a color-group mixing angle $\theta\approx 30^\circ$ while the second is only true for $\theta=0^\circ$. }  Second, requiring the $SU(3)$ gauge couplings to remain below the critical value that would induce fermion condensation implies $\theta > 14^\circ$, which is slightly stronger than the criterion of ``perturbativity" adopted by \cite{Frampton:2009rk}; 
this would tend to rule out some areas of visible top asymmetry at larger axigluon masses.   A similar bound is obtained by requiring the axigluon's width to be no larger than its mass.  Precision electroweak constraints from $\Delta\rho$ and $Z\bar{b}_L b_L$ in this model are relatively mild and do not impact the region of large $A^t_{FB}$.  Our results suggest that the axigluon model cannot produce as large an enhancement of $A^t_{FB}$ as previously supposed, while remaining consistent with other data.  Moreover,  we find that the $A^t_{FB}$ distributions as a function of $M_{t\bar{t}}$-edge do not obviously resemble the CDF data. 

In this context, it is useful to consider whether an extension toward an extra dimensional model in the continuum limit would enhance the size of the axigluon effect, for a given axigluon mass.  Towards this end, we consider extending the gauge symmetry to $SU(3) \times SU(3) \times SU(3)$.   We demonstrate that such an extension, which would be the first step in ``un-blocking" or ``un-deconstructing"   \cite{ArkaniHamed:2001ca,Hill:2000mu} the model towards a five-dimensional $SU(3)$ gauge theory, can only {\it dilute} the size of the axigluon effect on the top forward-backward asymmetry, regardless of how the quarks are charged under the various $SU(3)$ gauge groups.

We conclude that axigluon models (or their coloron \cite{Chivukula:1996yr} and topgluon \cite{Hill:1991at} cousins) are unlikely to be the source of the observed top quark asymmetry.

%%%%%%%%%%%%%%%%%%%
\section{The Model}

\subsection{Gauge Sector}

\begin{figure}[b]
\includegraphics[width=2.5in]{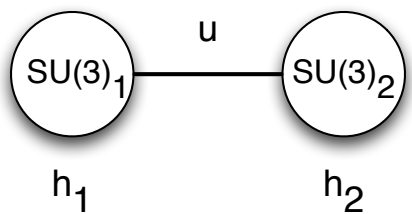}
\caption{Sketch of the color sector of the axigluon model in Moose notation  \cite{Georgi:1985hf}, showing the two gauge groups $SU(3)_i$, their associated gauge couplings $h_i$, and the condensate that breaks the color symmetries to their diagonal subgroup.}
\label{fig:twosite}
\end{figure}

We will describe the essential features of the model here; further details are given in \cite{Frampton:2009rk}. At high energies, the color sector of the model has an $SU(3)_{1} \otimes SU(3)_{2}$ gauge symmetry, with gauge couplings  $h_1$ and $h_2$; the electroweak gauge sector is as in the standard model.  The extended color group breaks to its diagonal subgroup, which we associate with $SU(3)_{QCD}$, when a Higgs field transforming  as a $\left(3,\bar{3}\right)$ acquires a diagonal vev of strength $u$.  The color sector of the model is summarized in Fig. \ref{fig:twosite} in Moose notation \cite{Georgi:1985hf}. The mass-squared matrix for the colored gauge-bosons is then
%%%%
\begin{equation}
\frac{u^2}{4}   
\left(
 \begin{array}{cc}
h_1^2&-h_1h_2\\
-h_1h_2&h_2^2\\
\end{array}
\right)~.
\label{MassMatrix}
\end{equation}
%%%
Defining $\sin{\theta} \equiv h_1/\sqrt{h_1^2 + h_2^2}$ and $\cos{\theta} \equiv h_2/\sqrt{h_1^2 + h_2^2}$, we obtain for the mass eigenstates $C_\mu^A$ (massive axigluons) and $G_\mu^A$ (massless fields identified with the QCD gluons), the relations
%%%
\begin{equation}
C^A_{\mu} = - \sin{\theta} \, A^A_{1 \mu} + \cos{\theta} \, A^A_{2 \mu} 
\qquad\qquad G^A_{\mu} =  \cos{\theta} \, A^A_{1 \mu} + \sin{\theta} \, A^A_{2 \mu} .
\label{MassEigenstates}
\end{equation}
%%%
The mass of the axigluon is
\begin{equation}
M_C \,=\,\frac{u}{\sqrt{2}} \sqrt{h_1^2 + h_2^2} ~,
\label{primeMassesC}
\end{equation}
%%%
and the coupling strength of the gluon ($g_S$) is given by
%%%%
\begin{equation}
g_S \equiv \frac{h_1 h_2}{\sqrt{h_1^2+h_2^2}} = h_1\cos\theta = h_2 \sin\theta ~.
\label{eq:gcouplS}
\end{equation}
%%%
The axigluon and gluon couple, respectively, to 
\begin{equation}
g_S J^\mu_C \equiv g_S (- J^\mu_1 \tan\theta + J^\mu_2 \cot\theta) \qquad\qquad   g_S J^\mu_G \equiv g_S (J^\mu_1 + J^\mu_2)\,,
\end{equation}
where $J^\mu_i$ is the current of quarks charged under color group $SU(3)_i$.   At energies below the axigluon mass, axigluon exchange induces the following four-fermion interaction among quarks:
\begin{equation}
{\cal L}_{FF}^{2} = - \frac{g_S^2}{2 M_C^2} J^\mu_C J_{C\,\mu}\,.
\end{equation}
Applying Eqs. (\ref{primeMassesC}) and (\ref{eq:gcouplS}) reveals the alternative, and also useful, form
\begin{equation}
{\cal L}_{FF}^{2} = - \frac{1}{u^2}(J^\mu_2 - \cos^2\theta J^\mu_G)^2 \, .
\label{eq:Lff2}
\end{equation}

\subsection{Quark Charge Assignments}

Our discussion of fermion charge assignments employs the weak gauge eigenstate fermions.  To avoid tree-level flavor-changing neutral currents (FCNC) in the first and second generations, we assume that the $u$, $d$, $c$, and $s$ quarks have the same color charges, and denote them by $q$; FCNC for the third generation are discussed in Section III.  Because the model is intended to explain the measured value of $A^t_{FB}$, the quark charge assignments under the color groups must enable the axigluon coupling to the fermions to satisfy $g^t_A g^q_A < 0$, the condition required \cite{Ferrario:2009bz} to increase $A^t_{FB}$ with respect to the standard model value without significantly altering the invariant top-pair mass.

\begin{table}[tb]
\caption{Distinct patterns of possible quark color assignments; all others are equivalent via exchange of the color groups.  A quark listed under a given group transforms as a fundamental under that group; if not listed, it is a singlet under that group. Only pattern 5 can lead to an enhancement of $A^t_{FB}$.}
\begin{center}
\begin{tabular}{|c||ccc|cccc|} \hline
$\phantom{\dfrac{\strut}{\strut}}$ && $ SU(3)_1 $&&&  $\ \ \ \ \ SU(3)_2$&& \\  \hline\hline
$\phantom{\dfrac{\strut}{\strut}} $ Pattern 1$\ \  $&&&  & $\ \ \ (t,b)_L,\ \ \ $&$\ \ \ q_L, \ \ \ $&$\ \ \ t_R, b_R,\ \ \ $&$\ \ \ q_R\ \ \ $ \\  \hline 
$\phantom{\dfrac{\strut}{\strut}} $ Pattern 2$\ \  $&&& $\ \ \ q_R\ \ \ $ & $\ \ \ (t,b)_L,\ \ \ $&$\ \ \ q_L, \ \ \ $&$\ \ \ t_R, b_R\ \ \ $& \\  \hline 
$\phantom{\dfrac{\strut}{\strut}} $ Pattern 3$\ \  $&&$\ \ \ t_R, b_R\ \ \ $&  & $\ \ \ (t,b)_L,\ \ \ $&$\ \ \ q_L, \ \ \ $&&$\ \ \ q_R\ \ \ $ \\  \hline 
$\phantom{\dfrac{\strut}{\strut}} $ Pattern 4$\ \  $&$\ \ \ q_L \ \ \ $&&  & $\ \ \ (t,b)_L,\ \ \ $&&$\ \ \ t_R, b_R,\ \ \ $&$\ \ \ q_R\ \ \ $ \\  \hline 
$\phantom{\dfrac{\strut}{\strut}} $ {\bf \large Pattern 5}$\ \  $&$\ \ \ ${\large\bf q}$_{\bf L}, \ \ \ $&$\ \ \ ${\large\bf t}$_{\bf R}$, {\large \bf b}$_{\bf R}\ \ \ $  && \ \ \ {\large\bf  (t,b)}$_{\bf L},\ \ \ $&&&\ \ \ {\large\bf q}$_{\bf R}\ \ \ $ \\  \hline 
$\phantom{\dfrac{\strut}{\strut}} $ Pattern 6$\ \  $&$\ \ \ q_L,\ \ \ $&& $\ \ \ q_R\ \ \ $  & $\ \ \ (t,b)_L,\ \ \ $&&$\ \ \ t_R, b_R\ \ \ $&\\  \hline 
$\phantom{\dfrac{\strut}{\strut}} $ Pattern 7$\ \  $&&$\ \ \ t_R, b_R,\ \ \ $& $\ \ \ q_R\ \ \ $ & $\ \ \ (t,b)_L,\ \ \ $&$\ \ \ q_L \ \ \ $&& \\  \hline 
$\phantom{\dfrac{\strut}{\strut}} $ Pattern 8$\ \  $&$\ \ \ q_L, \ \ \ $&$\ \ \ t_R, b_R,\ \ \ $&$\ \ \ q_R\ \ \ $ & $\ \ \ (t,b)_L\ \ \ $&&&\\  \hline 
\end{tabular}
\end{center}
\label{tab:one}
\end{table}

Since there are two gauge groups and four sets of SM quarks ($q_L$; $\ q_R$; $\ (t,b)_L$; $\ t_R$, $b_R$) there are only eight distinct patterns for assigning the fermion charges\footnote{One might, alternatively, assign $b_R$ to transform like the $q_R$, but it would not materially affect the outcome.};
all others are equivalent to these via exchange of $SU(3)_1 \leftrightarrow SU(3)_2$.  These patterns are shown in Table~\ref{tab:one}. 

Any assignment for which the two chirality components $f_L$ and $f_R$ of a fermion $f$ transform under the same $SU(3)$ is vectorial, ensuring that $g^f_A = 0$; this applies to patterns 1-4, 6, and 8, so that none of these assignment schemes is relevant for enhancing $A^t_{FB}$.  Flavor-universal coloron models \cite{Chivukula:1996yr} conform to pattern 1, while topcolor models \cite{Hill:1991at} fall under pattern 6.   While pattern 7 does not yield a purely vectorial coupling for any fermion, it is flavor-universal, so $g^t_A = g^q_A$; this is the traditional axigluon \cite{Frampton:1987dn} charge assignment. As discussed in \cite{Sehgal:1987wi,Choudhury:2007ux,Antunano:2007da}, for the classic axigluon model where $g_1 = g_2$, the value of $A^t_{FB}$ can only be more negative than in the standard model; if $g_1 \neq g_2$, the squared-amplitude from axigluon exchange can generate a modest  positive value of $A^t_{FB}$ \cite{Martynov:2009en,Martynov:2010ed} but only at the expense of a significant alteration of the invariant top-pair mass \cite{Ferrario:2009bz}.  The only pattern of possible interest is pattern 5; this is one chosen by \cite{Frampton:2009rk} and the one we investigate here. From the form of the four-fermion interaction (\ref{eq:Lff2}) one can see immediately
\begin{equation}
g^t_L,\  g^q_R\  \propto\  1 - \cos^2\theta \qquad\qquad  g^t_R,\ g^q_L\  \propto\  - \cos^2\theta \, ,
\end{equation}
so that $g^t_A g^q_A < 0$ is satisfied.  More specifically, one finds that the vector and axial couplings of the axigluon to the fermons are \cite{Frampton:2009rk}:
\begin{equation}
g^t_V = g^q_V = - g_S \cot{2\theta}  \qquad\qquad g^t_A = - g^q_A = g_S \csc{2\theta} \, .
\label{eq:couplingss}
\end{equation}
Spectator fermions will be needed to cancel gauge anomalies, and \cite{Frampton:2009rk} employs a fourth generation of quarks for this purpose. The Yukawa couplings for the quarks (but not the leptons) will need to be modified because the different quark chiralities are charged under different $SU(3)$ groups. Finally, while the full range of $\theta$ is from $0$ to $90^\circ$, the model's phenomenology is symmetric under the exchange $\theta \to 90^\circ - \theta$.  

%%%%%%%%%%%%%%%%%%%
\section{Low-energy Phenomenology}

\begin{figure}[tb]
\includegraphics[width=4in]{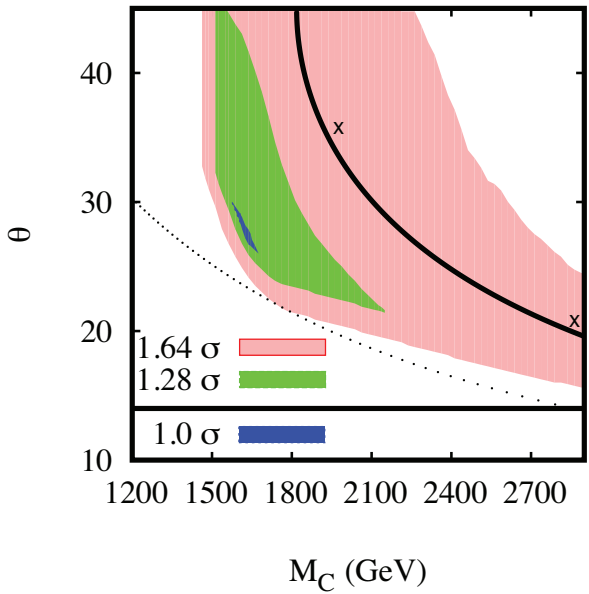}
\caption{Figure originally from \cite{Frampton:2009rk} showing the $M_C$ v.s. $\theta$ plane, with our new phenomenological bounds overlaid (regions above the black curves are allowed).  Angle $\theta$ is shown in units of degrees; since the model has a [$\theta \leftrightarrow 90^\circ - \theta$] symmetry, only angles up to $45^\circ$ are shown.  The shaded regions, calculated in \cite{Frampton:2009rk},  indicate where the predicted value of $A^t_{FB}$ in the axigluon model agrees with data to within $1\sigma$ (small dark [blue] shaded region), $1.28\sigma$ (medium [green] shaded region), or $1.64\sigma$ (large pale [pink] shaded region).   We have added curves showing our bounds from $B_d$-meson mixing (solid black curve, 95\% CL),  $\Delta\rho$ (dotted black curve, 95\%CL), and fermion condensation (horizontal black line);  the regions above our curves/lines are allowed. Note that our limit from $B_d$-meson mixing excludes the region where $A^t_{FB}$ showed the greatest agreement with the data .  The two crosses show the approximate locations of the sample points for the distributions discussed in section III.E.}
\label{fig:one}
\end{figure}

The gauge sector of the model includes two more parameters than the standard model gauge sector (a gauge coupling and a vacuum expectation value), so the phenomenology associated with the new physics may be summarized within the $M_C$ v.s. $\theta$ plane.   Fig.~\ref{fig:one} contains such a summary; it starts with a figure from \cite{Frampton:2009rk} indicating shaded regions of relatively large $A^t_{FB}$ and overlays the new phenomenological bounds we derive in this section (regions above the black curves are allowed).  Angle $\theta$ is shown in units of degrees; since the model has a [$\theta \leftrightarrow 90^\circ - \theta$] symmetry, only angles up to $45^\circ$ are shown.  The shaded regions, calculated in \cite{Frampton:2009rk},  indicate where the predicted value of $A^t_{FB}$ in the axigluon model agrees with data\footnote{Agreement with the top pair production cross-section was also required in \cite{Frampton:2009rk}.}  to within $1\sigma$ (small dark blue region, $\sim 68\%$CL), $1.28\sigma$ (medium green region, $\sim 80\%$CL), or $1.64\sigma$ (large pale pink region $\sim 90\%$CL).   We have added curves showing our bounds from $B_d$-meson mixing (solid black curve, 95\%CL),  $\Delta\rho$ (dotted black curve, 95\%CL), and fermion condensation (horizontal black line);  the regions above our curves/lines are allowed. Note that our limit from $B_d$-meson mixing excludes the region where $A^t_{FB}$ showed the greatest agreement with the data.  The two crosses show the approximate locations of the sample points for the distributions discussed in section III.E.

\subsection{Neutral $B_d$-meson Mixing}

The flavor-changing four-fermion interactions induced by axigluon exchange will alter the predicted rate of $B_d$-meson mixing.  Since the fermion charge assignments discussed above are for the weak gauge eigenstate fields, which are related to the mass eigenstate fields through the CKM matrix $V_{ij}$, the operator of particular concern in the axigluon model is (adapting the analysis of \cite{Braam:2007pm}):
\begin{equation}
\frac{8 \pi \alpha_s}{3 M_c^2 \sin^2(2\theta)} (V_{td}^* V_{tb})^2 (\bar{b}_L \gamma^\mu d_L) (\bar{b}_L \gamma_\mu d_L)\,.
\end{equation}
The UTFit Collaboration has derived  \cite{Bona:2007vi} constraints on general $\Delta F = 2$ four-fermion operators that affect neutral meson mixing, including the effects of running from the new physics scale down to the meson scale and interpolating between quark and meson degrees of freedom.   For a model like the axigluon model, with tree-level FCNC and a flavor structure like that of the SM, Ref. \cite{Bona:2007vi} writes the applicable operator as
\begin{equation}
 \frac{(V_{td}^* V_{tb})^2}{\Lambda^2} (\bar{b}_L \gamma^\mu d_L) (\bar{b}_L \gamma_\mu d_L) \\
\end{equation}
and they obtain the 95\% CL limit $\Lambda > 1.8$ TeV.  Since $(8 \pi \alpha_s / 3) \approx 1$, the UTFit bound implies
\begin{equation}
M_c \sin(2\theta) > 1.8\ {\rm TeV}
\end{equation}
at 95\%CL which means that the region above the solid black curve in Fig. \ref{fig:one} is still allowed.  Clearly, this FCNC bound excludes the region where the value of $A^t_{FB}$ predicted by the axigluon model comes closest to agreement with the data.

\subsection{Fermion Condensation}

In the model proposed by \cite{Frampton:2009rk}, it is important that the couplings $h_i$ never become strong enough to cause fermions charged under group $SU(3)_i$ to condense.  Obtaining $g^t_V = -g^q_A \neq 0$ depends on having the $t_L$ and $t_R$ charged under different $SU(3)$ (so that top will have a non-zero axial coupling) and having $t_L$ and $q_R$ charged under the same $SU(3)$ (yielding the relative minus sign).  If the coupling of the $SU(3)$ group under which $t_L$ and $q_R$ are charged became strong, a $\bar{t}_L q_R$ condensate would form, effectively re-defining which right-handed state was part of the massive top quark and removing the prediction of a positive enhancement of $A^t_{FB}$.  To avoid this, we must ensure that $h_1$ and $h_2$ each lie below the critical value at which condensation would occur.  

At energies well below the mass of the axigluon, the self-consistent dynamical generation of masses for the fermions charged under $SU(3)_2$ occurs, in the Nambu--Jona-Lasinio  approximation \cite{Nambu:1961fr,Nambu:1961tp}, when the gap equation
%%%
\begin{equation}
m_f = \frac{3 m_f \alpha_s \cot^2\theta }{2\pi} \left[1-\left(\frac{m_f}{M_{C}}\right)^2 \ln{\left(\frac{M^2_{C}}{m_f^2}\right)}\right],
\label{mf}
\end{equation}
%%%
has a solution for $m_f > 0$. Accordingly we expect that condensation will not occur if $\alpha_s \cot^2\theta < 2 \pi / 3$.  A more complete analysis including QCD effects via the gauged NJL model  \cite{Appelquist:1988vi,Yamawaki:1988na,Kondo:1988qd,Miransky} yields the slightly stronger condition 
\begin{align}
\alpha_s\cot^2\theta <  \frac{2}{3} \pi - \frac{4}{3} \alpha_S 
\end{align}
This implies that one must have $\theta > 14^\circ$ to avoid condensation of the fermions charged under $SU(3)_2$ and, likewise $\theta < 76^\circ$ to avoid condensation of  fermions charged under group $SU(3)_1$.   These are stronger than the bounds on $\theta$ employed in \cite{Frampton:2009rk}; as indicated by Fig. \ref{fig:one}, for axigluon masses $M_c \laem 3$ TeV,  this bound on $\theta$ excludes some of the parameter space where the model predicts a $1.64\sigma$ agreement with the measured $A^t_{FB}$.

\subsection{Axigluon Width}

If one assumes, following the authors of  \cite{Frampton:2009rk}, that the heavy axigluon decays only to the standard model quarks, then its decay width is (ignoring corrections of order $(m_t / M_C)^2$):
\begin{equation}
\Gamma_C =  \frac{M_C}{24\pi} \cdot  6 \left[(g^t_V)^2 + (g^t_A)^2\right]
\end{equation}
where the factor of 6 arises because the magnitude of $g_V$ (or $g_A$) is the same for all quark flavors (see Eq. (\ref{eq:couplingss})). Hence, if one requires the axigluon to satisfy $ \Gamma _C/ M_C \laem 1$ so that it is clearly a distinct resonance, then $\theta$ is constrained to lie within essentially the same bounds as are required for avoiding fermion condensation ($12^\circ \laem \theta \laem 78^\circ$).

\subsection{Precision Electroweak Constraints}

Exchange of massive axigluons across the top and bottom quark loops in $W$ and $Z$ vacuum polarization diagrams alters the predicted value of $\Delta\rho$.   Updating the related limit \cite{Chivukula:1995dc,Chivukula:1996yr} on the effect of colorons with recent  experimental constraints \cite{Amsler:2008zzb} on $\Delta\rho$ yields the 95\%CL lower bound:
\begin{equation}
M_c > \cot\theta * 700\, {\rm GeV}
\end{equation}
The model parameter space where $A^t_{FB}$ is enhanced meets this constraint, as shown by the fact that it lies above the dotted black curve in Fig. \ref{fig:one}.

Because the third-generation fermions are treated differently than the light quarks, one should consider whether the $Z\bar{b}_L b_L$ coupling will be affected.  However, axigluon exchange across the $Z\bar{b}_L b_L$ vertex leads only to effects proportional to $m_b^2$ (because the resulting triangle diagram has no interior top quarks), and these are negligible.

\subsection{Asymmetry Distributions}

Having determined the region of parameter space that satisfies the
phenomenological bounds discussed above, we have generated plots of $A^t_{FB}$
as a function of the invariant top-pair mass $M_{t\bar{t}}$ and as a function of
the $M_{t\bar{t}}$ edge distribution.  Ref. \cite{Frampton:2009rk} presented
similar  plots for a sample point located at the region of greatest predicted
asymmetry ($M_C = 1525\, {\rm GeV}; \theta = 27^\circ$).  Since that point is
excluded by the data on $B_d\bar{B}_d$ mixing, we use two new sample points that
lie within the allowed region (at the 90\% CL): one at a relatively low mass and high angle ($M_C
= 2000\, {\rm GeV}; \theta = 35^\circ$) and the other at a higher mass and lower
angle ($M_C = 2850\, {\rm GeV}; \theta = 20^\circ$) as marked by crosses in 
Fig. \ref{fig:one}.

We follow the prescription given in Ref.~\cite{Frampton:2009rk} for calculating various distributions 
with CTEQ6L parton distribution functions~\cite{Pumplin:2002vw}, and both 
the factorization and renormalization scales are set to be equal to the 
top quark mass. As in Ref.~\cite{Frampton:2009rk}, the top quark mass is set to be 175 GeV 
and both the quark-antiquark ($q \bar q$) and gluon-gluon ($gg$) scattering processes are considered 
at the leading order. In the $q \bar q \rightarrow t \bar t$ channel, the complete gauge 
invariant set of Feynman diagrams, with gluon or axigluon propagator, is included, Since 
there is no direct triple coupling of $g-g-C$, the $gg \rightarrow t \bar t$  contribution is not 
modified at the leading order. 

\begin{figure}[b]
\vspace{.75cm}
\includegraphics[width=3.25in]{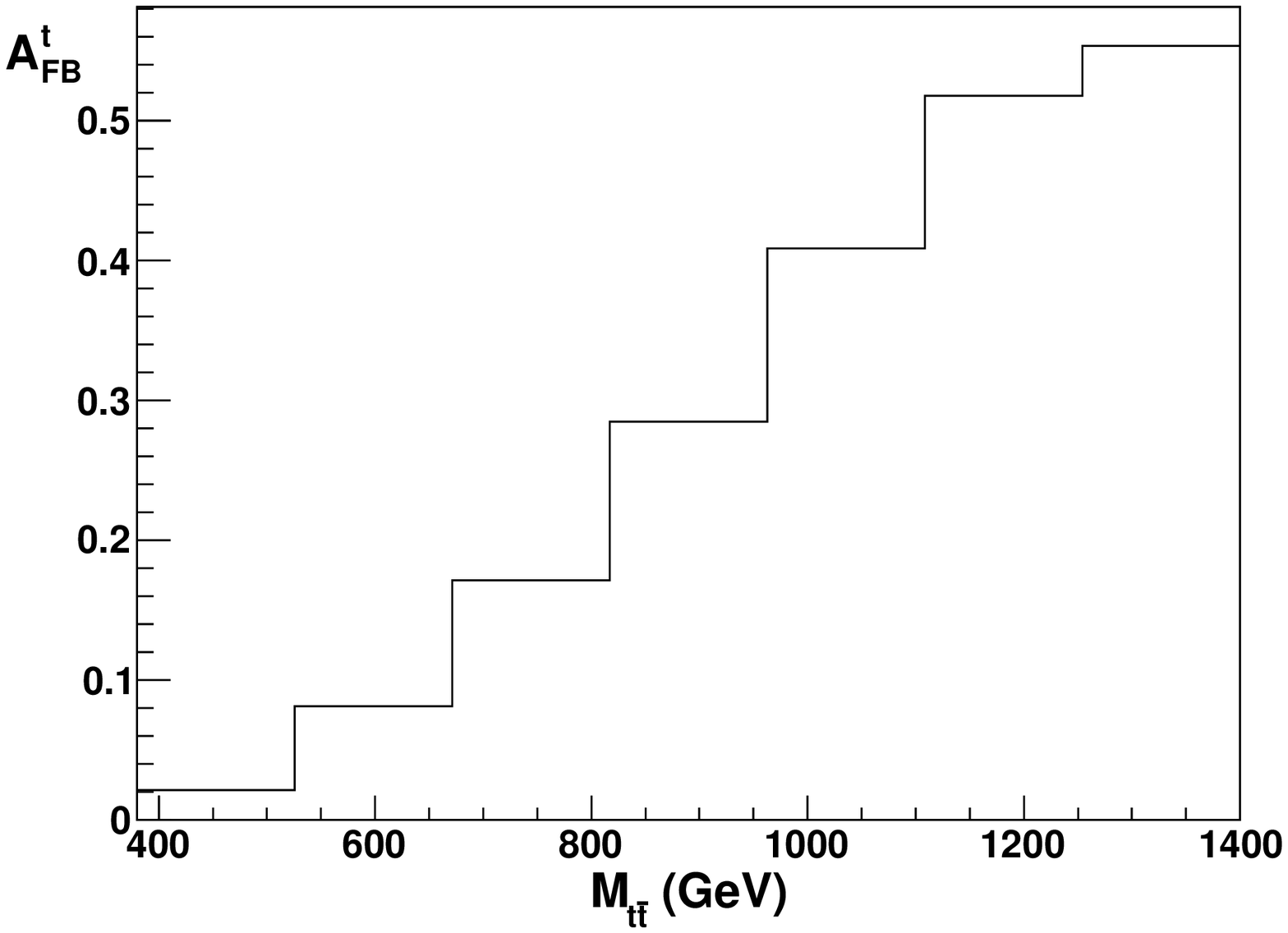} \includegraphics[width=3.6
in]{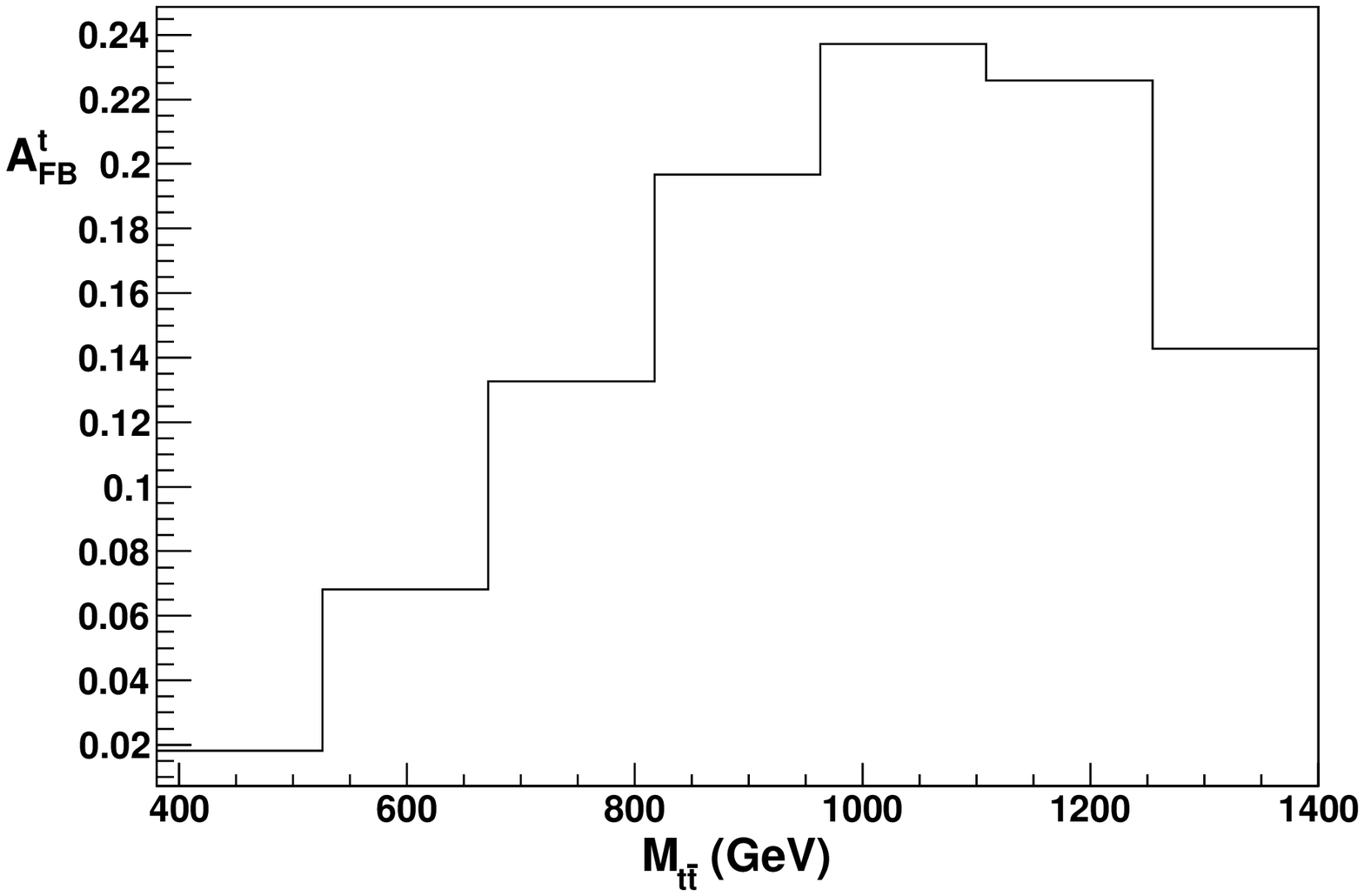}
\caption{Top-quark forward-backward asymmetry as a function of $M_{t\bar{t}}$ for our two sample points in the region allowed by $B_d$-meson mixing.  Left: $M_C = 2000 {\rm GeV}; \theta = 35^\circ$.  Right: $M_C = 2850 {\rm GeV}; \theta = 20^\circ$.  The peaked structure shown at right occurs only for low values of the mixing angle $\theta$.}
\label{fig:mtt}
\end{figure}

Fig.~\ref{fig:mtt} shows $A^t_{FB}$ as a function of $M_{t\bar{t}}$ for our two
sample points.  We find that the shape of this distribution is monotonically
increasing with invariant mass, unless theta is relatively low.  Hence, the
distinctive ``peaked" shape of this distribution discussed in 
\cite{Frampton:2009rk} in relation to their sample point is not a general
characteristic of the model in the allowed region.
Furthermore, the inclusive forward-backward asymmetry $A^t_{FB}$ is found to be $0.040$ and $0.032$ for $M_C=2000$ and $2850$, respectively. The corresponding forward-backward asymmetry defined in the center-of-mass frame of the $t \bar t$ system, via the polar angle of 
$t$ relative to the proton beam direction, is  $0.055$ and $0.044$, respectively.

\begin{figure}[t]
\includegraphics[width=3.5in]{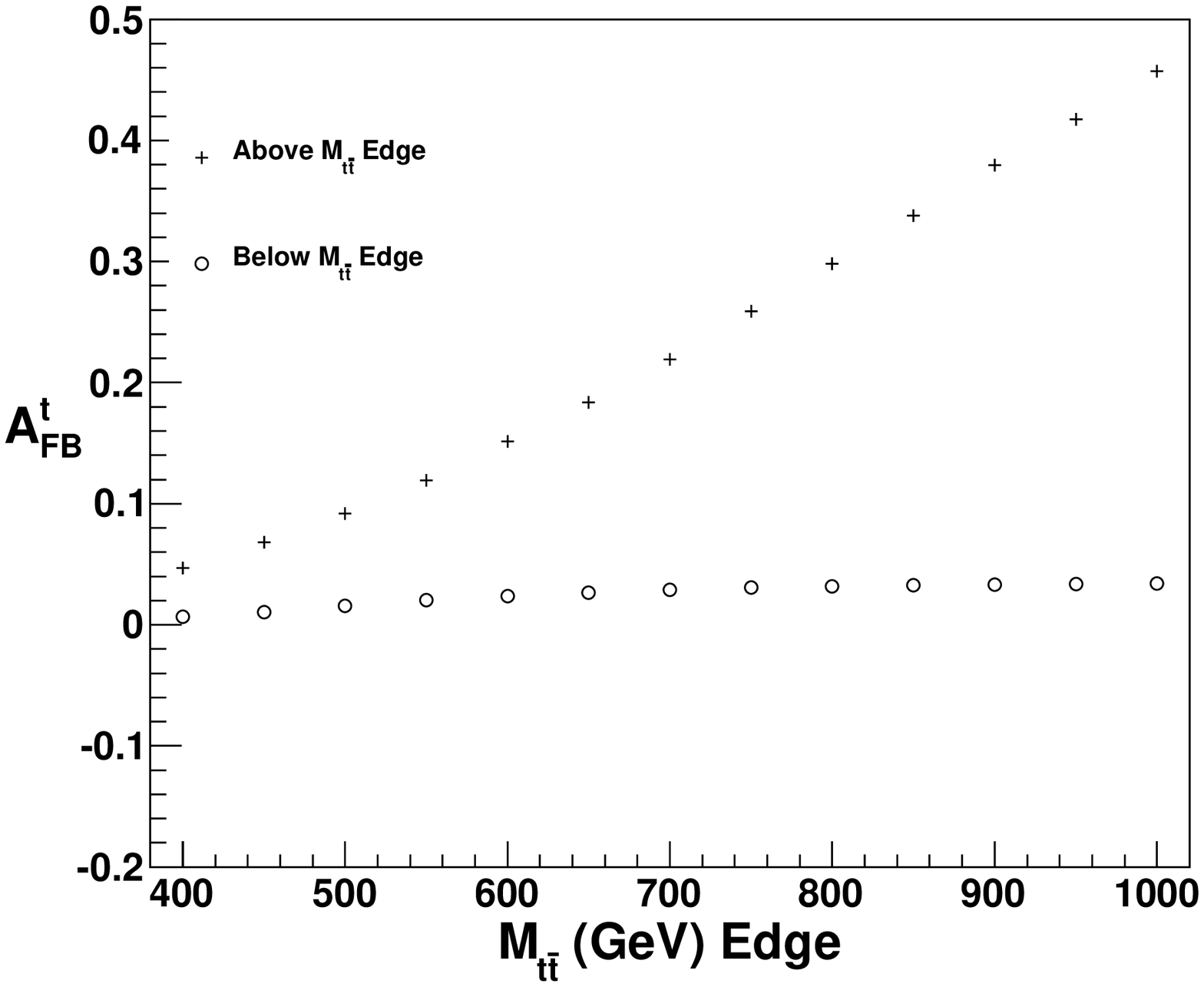} \includegraphics[width=3.25
in]{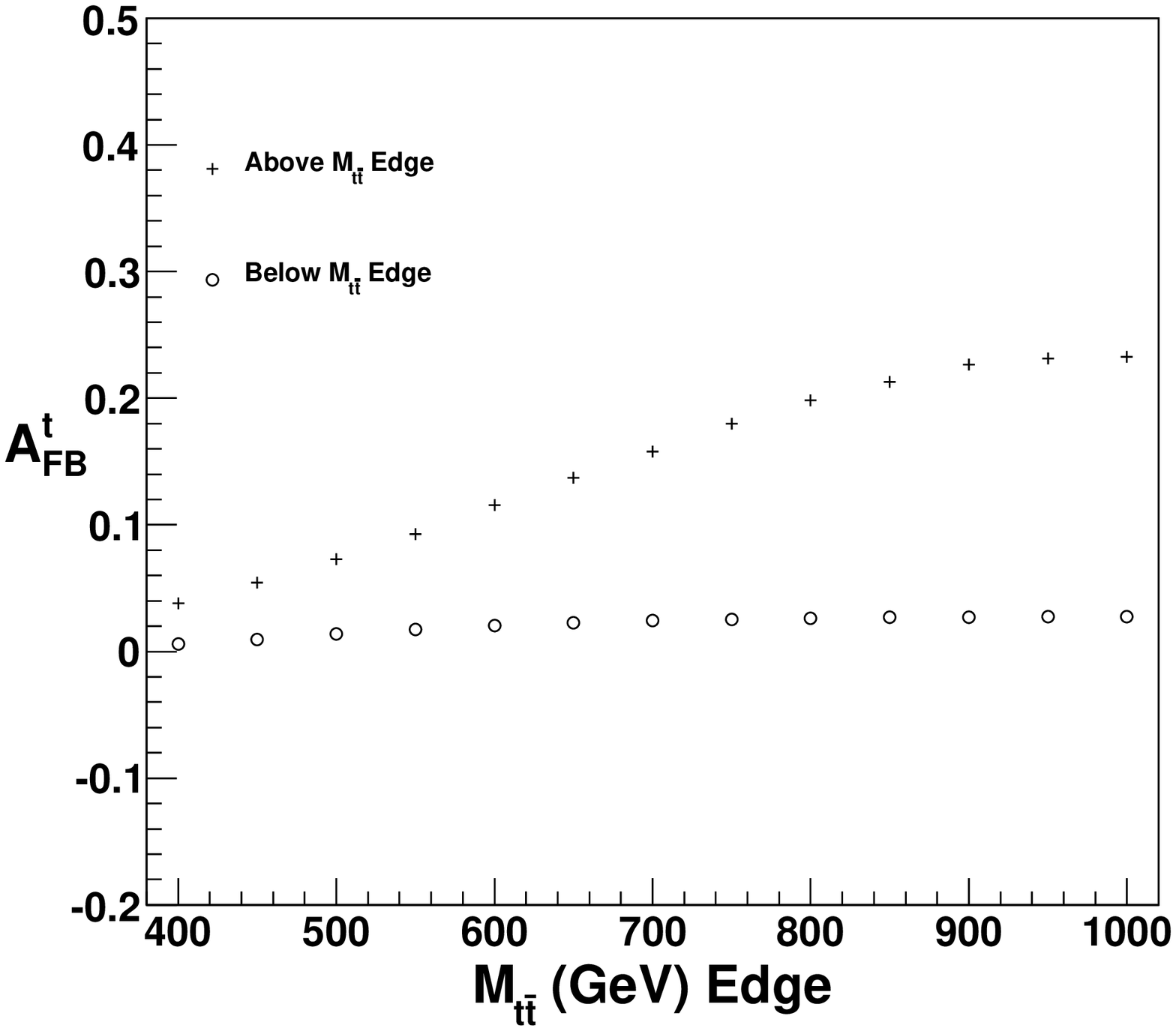}
\caption{Top-quark forward-backward asymmetry for events above (or below) the shown $M_{t\bar{t}}$ value for our two sample points in the region allowed by $B_d$-meson mixing.  Left: $M_C = 2000 {\rm GeV}; \theta = 35^\circ$.  Right: $M_C = 2850 {\rm GeV}; \theta = 20^\circ$.  Neither of the ``below-edge'' curves resemble the data \cite{CDF-public}; nor does the``above-edge'' curve at left; the ``above-edge'' curve at right might arguably be consistent with the data within its large error-bars, but is not obviously of the same shape.}
\label{fig:mttedge}
\vspace{.5cm}
\end{figure}

Fig.~\ref{fig:mttedge} shows $A^t_{FB}$ for the top pair events above (or below)
a given $M_{t\bar{t}}$ threshold; again, one distribution is shown for each of
our sample points.  Ref. \cite{Frampton:2009rk} noted that such an
$M_{t\bar{t}}$-edge distribution can be directly compared with the CDF data
\cite{CDF-public}.  More precisely, our model curves show the new-physics
contribution to $A^t_{FB}$; one can compare this with the difference between the
data and the NLO QCD theory curve shown in \cite{CDF-public}.  We see that
neither curve in the left-hand plot resembles the difference between data and
standard model theory; nor does the ``below'' curve in the right-hand plot; the
``above'' curve in the right-hand plot might arguably be consistent with the
data within its large error-bars, but is not obviously of the same shape as the
difference between data and standard model theory.

%%%%%%%%%%%%%%%%%%%
\section{Extension To A Larger Color Group}

\subsection{An $SU(3)^3$ Model}

We now consider whether an extension toward an extra dimensional model in the continuum limit would enhance the size of the axigluon effect on the top asymmetry, for a given axigluon mass.  Towards this end, we consider extending the gauge symmetry to $SU(3)^3$.   Such an extension would be the first step in ``un-blocking" or ``un-deconstructing"   \cite{ArkaniHamed:2001ca,Hill:2000mu} the model towards a a five-dimensional $SU(3)$ gauge theory.

\begin{figure}[t]
\includegraphics[width=3in]{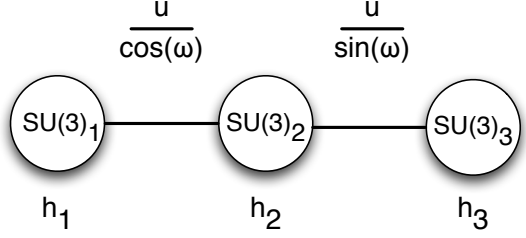}
\caption{Sketch of the extended axigluon model in Moose notation, showing the three gauge groups $SU(3)_i$, their associated gauge couplings $h_i$, and the two condensates that break pairs of color symmetries to their diagonal subgroups.}
\label{fig:threesite}
\end{figure}

Let us consider an $SU(3)_1 \times SU(3)_2 \times SU(3)_3$ extended color group, as shown in Fig. \ref{fig:threesite} with associated gauge couplings
\begin{equation}
h_1 = \frac{g_s}{\sin\theta} \qquad\qquad h_2 = \frac{g_s}{\cos\theta\sin\phi} \qquad\qquad h_3 = \frac{g_s}{\cos\theta\cos\phi}\,,
\end{equation}
and fermion currents $J^\mu_i$.  We assume that symmetry breaking now proceeds via the expectation values of two effective Higgs fields in bi-fundamental color representations: 
\begin{eqnarray}
\Phi_a {\rm\  transforms\  as\  a\ } (3, \bar{3}, 1) & {\rm and\  its\  vev\  is\ } \langle\Phi_a\rangle = \frac{u}{\cos\omega} \cal{I}\\
\Phi_b {\rm \ transforms\  as\  a\ } (1, 3, \bar{3}) & {\rm and\  its\ vev\  is\ } \langle\Phi_b\rangle = \frac{u}{\sin\omega} \cal{I}\, .
\end{eqnarray}
When the fields acquire their vacuum expectations values, $\langle\Phi_a\rangle$ breaks $SU(3)_1\times SU(3)_2$ down to the diagonal subgroup $SU(3)_{1+2}$, while $\langle\Phi_b\rangle$ breaks $SU(3)_2\times SU(3)_3$ down to the diagonal subgroup $SU(3)_{2+3}$; with this pattern, $SU(3)_C \equiv SU(3)_{1+2+3}$ remains unbroken.

The mass matrix for the gauge bosons may be written as:
\begin{equation}
{\cal M}^2= \frac{g_S^2 u^2}{4} \begin{pmatrix}
\frac{1}{\sin^2\theta \, \cos^2\omega} & -\frac{1}{\sin\theta\cos\theta\sin\phi\,\cos^2\omega} & 0 \\
 -\frac{1}{\sin\theta\cos\theta\sin\phi\,\cos^2\omega}  & \frac{1}{\cos^2\theta \sin^2\phi\,\cos^2\omega \sin^2\omega} & -\frac{1}{\cos^2\theta \sin\phi \cos\phi\,\sin^2\omega}\\
 0 & -\frac{1}{\cos^2\theta \sin\phi \cos\phi\,\sin^2\omega} & \frac{1}{\cos^2\theta \cos^2\phi\,\sin^2\omega}
\end{pmatrix} \, .
\label{eq:mmatr}
\end{equation}
To analyze this system, it is convenient to change to the basis\footnote{This choice
is motivated by the analysis in \protect\cite{Georgi:1989xz}.}
\begin{align}
G^\mu & = \sin \theta A^\mu_1 + \cos\theta\sin\phi A^\mu_2 + \cos\theta A^\mu_3\\
C^\mu_a & = -\cos\theta A^\mu_1 + \sin\theta \sin\phi A^\mu_2 + \sin\theta \cos\phi A^\mu_3\\
C^\mu_b & = -\cos\phi A^\mu_2 + \sin\phi A^\mu_3~,
\end{align}
where $A^\mu_{1,2,3}$ are the gauge bosons of the groups $SU(3)_{1,2,3}$ respectively. This basis
is convenient because $G^\mu$ is the massless eigenstate field associated with the
unbroken color group $SU(3)_C$, and couples to fermion currents
\begin{equation}
g_S J^\mu_G \equiv g_S ( J^\mu_1 + J^\mu_2 + J^\mu_3)\,,
\end{equation}
just like the gluon.  The fields $C^\mu_{a,b}$ couple to the currents
\begin{eqnarray}
g_S {J}_a^\mu &= \frac{g_S}{\cos\theta \sin\theta}\left(J^\mu_2+J^\mu_3-\cos^2\theta J^\mu_G\right)\\
g_S {J}^\mu_b &= \frac{g_S}{\cos\theta \cos\phi \sin\phi}\left(\sin^2\phi J^\mu_3-\cos^2\phi J^\mu_2\right)~.
\end{eqnarray}

At low energies, exchange of the massive colored bosons gives rise to four-fermion interactions of the form
\begin{equation}
{\cal L}_{FF}^3 = - \frac{g_S^2}{4} \begin{pmatrix} J^\mu_a\,, & J^\mu_b \end{pmatrix}
\left( {\cal M}^2_{2\times 2} \right)^{-1}
 \begin{pmatrix} J^\mu_a\\ J^\mu_b \end{pmatrix} \,,
 \end{equation}
 where ${\cal M}^2_{2 \times2} $ is the portion of the mass matrix (\ref{eq:mmatr}) restricted  to the 2-dimensional space spanned by $C^\mu_{a,b}$,
\begin{equation}
{\cal M}^2_{2\times 2}=
\frac{u^2 g_S^2 }{4 \cos^2\theta \cos^2\omega}
\begin{pmatrix}
 \frac{1}{\sin^2\theta} & -\frac{1}{\sin^2\theta} \frac{\cos^\phi}{\sin\phi}  \\
-\frac{1}{\sin^2\theta} \frac{\cos^\phi}{\sin\phi} \ \ \ \ \ \ \ \ & (\frac{1}{\cos^2\phi\sin^2\phi}\frac{\cos^2\omega}{\sin^2\omega} + \frac{\cos^2\phi}{\sin^2\phi})
\end{pmatrix}~.
\end{equation}
After some simplification, the four-fermion interactions emerge in the following streamilined form:
\begin{equation}
{\cal L}_{FF}^3 = - \frac{\cos^2\omega}{u^2} \left(
J^\mu_2+J^\mu_3-\cos^2\theta J^\mu_G\right)^2 -
\frac{\sin^2\omega}{u^2} \left(
J^\mu_3-\cos^2\phi \cos^2\theta J^\mu_G\right)^2~.
\label{eq:l3ff}
\end{equation}

\subsection{Quark Charge Assignments and $A^t_{FB}$}

Our underlying question is whether the quark charges under the $SU(3)^3$ gauge group can be assigned so as to increase  $A^t_{FB}$ relative to the prediction in the $SU(3)^2$ axigluon model.   To help us think about this, we should recall that the enhancement arises when the four-fermion operators of Eq. (\ref{eq:l3ff}) interfere with the QCD amplitude for top-pair production; only when the four-fermion operators contain terms proportional to $g^t_A g^q_A < 0$ does the predicted value of $A^t_{FB}$ increase.

First, consider the situation where $SU(3)_2$ is fermiophobic. In this case, Eq. (\ref{eq:l3ff}) reduces to
\begin{eqnarray}
{\cal L}_{FF}^3  &\rightarrow & - \frac{\cos^2\omega}{u^2} \left(J^\mu_3-\cos^2\theta J^\mu_G\right)^2 -\frac{\sin^2\omega}{u^2} \left(
J^\mu_3-\cos^2\phi \cos^2\theta J^\mu_G\right)^2\\
&=& - \frac{1}{u^2} \left(J^\mu_3-\cos^2\theta J^\mu_G\right)^2 - \frac{2 \sin^2\omega}{u^2} \left(\cos^2\theta\sin^2\phi J^\mu_G \right) \left(J^3_\mu - \cos^2\theta J^\mu_G\right) - \frac{\sin^2\omega}{u^2}\left(\cos^2\theta\sin^2\phi J^\mu_G\right)^2
\label{eq:l3ffrewrite}
\end{eqnarray}
Once the four-fermion operators are re-written as in Eq. (\ref{eq:l3ffrewrite}), it is clear that the last term is purely vectorial and the middle is of the form $g^f_V g^{f'}$; neither term has the $g^t_A g^q_A$ form required to contribute to $A^t_{FB}$.  The first term looks familiar: it is effectively identical to the four-fermion operator in the $SU(3)^2$ model from Eq. (\ref{eq:Lff2}).  Moreover, if all of the fermions are charged under two color groups, then as argued earlier, the only way to get a non-zero contribution to the asymmetry is to use an equivalent of the ``pattern 5" charge assignment, as shown in Table \ref{tab:two}. 
\begin{table}[t]
\caption{This is the only pattern of quark charge assignments under the color groups that both renders $SU(3)_2$ fermiophobic and yields a non-zero contribution to the top asymmetry.  A quark listed under a given group transforms as a fundamental under that group; if not listed, it is a singlet under that group.}
\begin{center}
\begin{tabular}{|c||c|c|c|} \hline
$\phantom{\dfrac{\strut}{\strut}}$ & $ SU(3)_1 $&  $\ \ \ SU(3)_2\ \ \ $ & $SU(3)_3$ \\  \hline\hline
$\phantom{\dfrac{\strut}{\strut}} $ Pattern 5$^\prime\ \  $&  $\ \ \ q_L, \ \ \ t_R, b_R \ \ \ \ \ \ \ \ \ \ \ \ $&  & $\ \ (t,b)_L,\ \ \ \ \ \ \ \ \ \ \ \ \ \ \ \ \ \ \ \ \ \ q_R \ \ \  $\\  \hline 
\end{tabular}
\end{center}
\label{tab:two}
\end{table}
We conclude that when the fermions transform only under the ``outer" two groups in the linear moose, the effect on $A^t_{FB}$ is precisely as in the axigluon model with only two color groups.

\begin{table}[b]
\caption{Patterns of possible quark charge assignments under the color groups for either $q$ or $t$ when $SU(3)_2$ is not fermiophobic.  A quark listed under a given group transforms as a fundamental under that group; if not listed, it is a singlet under that group.}
\begin{center}
\begin{tabular}{|c||c|c|c|} \hline
$\phantom{\dfrac{\strut}{\strut}}$ & $ \ \ \ SU(3)_1\ \ \  $&  $\ \ \ SU(3)_2\ \ \ $ & $\ \ \ SU(3)_3\ \ \ $ \\  \hline\hline
$\phantom{\dfrac{\strut}{\strut}} $ Pattern A$\ \  $& $f_R$ & $f_L$ & \\  \hline 
$\phantom{\dfrac{\strut}{\strut}} $ Pattern B$\ \  $&  & $f_L, f_R$ & \\  \hline 
$\phantom{\dfrac{\strut}{\strut}} $ Pattern C$\ \  $&  & $f_L$ & $f_R$\\  \hline 
\end{tabular}
\end{center}
\label{tab:three}
\end{table}

Now consider the effect of allowing a fermion to be charged under group $SU(3)_2$.   Its chiral partner must be charged under one of the three $SU(3)$ groups, giving rise to three possible patterns as shown in Table \ref{tab:three} (exchanging $L \leftrightarrow R$ makes no difference).If one chooses pattern A, note that the second four-fermion operator in ${\cal L}_{FF}^3$ (see Eq.  (\ref{eq:l3ff})) treats fermions charged under groups 1 and 2 identically and does not include an axial coupling for the fermion; only the first operator can contribute to $A^t_{FB}$, and it has a suppression factor of $\cos^2\omega$ relative to ${\cal L}^2_{FF}$ (see Eq. (\ref{eq:Lff2})).  If one selects pattern B, the fermion's couplings are entirely vectorial, $g^f_A = 0$.   If one elects pattern C, we have the opposite situation to pattern A: the first operator in ${\cal L}_{FF}^3$ treats fermions charged under groups 2 and 3 the same, so only the second operator, which is suppressed by $\sin^2\omega$ contributes to $A^t_{FB}$.  Since the top asymmetry depends on the product on the top and light-quark axial charges, this argument applies equally well for $t$ and $q$.  In other words, in the $SU(3)^3$ having any fermion charged under the middle group of the linear moose instead of the outer groups leads to a reduction of $A^t_{FB}$, relative to the $SU(2)^2$ case.  

Extending this to an extra-dimensional language, one would say that localizing the quarks on the branes leads to the maximum enhancement of the top asymmetry; allowing fermions to delocalize into the bulk can only reduce the effect.  Hence, extending the $SU(3)^2$ model of  \cite{Frampton:2009rk} into an extra-dimensional framework will not increase the predicted value of the top asymmetry.

%%%%%%%%%%%%%%%%%%%
\section{Conclusions}

We have studied the axigluon model  proposed by Frampton, Shu, and Wang  \cite{Frampton:2009rk}  in response to the experimentally observed enhancement of the forward-backward asymmetry in top quark production.  We find that limits from data on $B_d$ mixing exclude the region in axigluon mass vs. gauge mixing parameter space in which the enhancement of $A^t_{FB}$ comes closest to agreement with experiment, and that bounds from the requirement that the gauge couplings be weak enough to prevent fermion condensation rule out a further region of larger axigluon masses.  Our results suggest that the axigluon model cannot produce even as large an enhancement of $A^t_{FB}$ as previously supposed and that the $A^t_{FB}$ distributions as a function of $M_{t\bar{t}}$-edge do not obviously resemble the CDF data. 

We have seen that the pattern of quark charge assignments chosen by \cite{Frampton:2009rk} is the only one for which a positive enhancement of $A^t_{FB}$ occurs in an $SU(3)^2$ model; other coloron, axigluon, or topcolor models based on $SU(3)^2$ gauge groups give null or wrong-sign effects.  Moreover, we have demonstrated that extending the color sector to $SU(3)^3$, which would be the first step in ``un-deconstructing" the model towards a a five-dimensional $SU(3)$ gauge theory, can only {\it dilute} the size of the axigluon effect on the top forward-backward asymmetry, regardless of how the quarks are charged under the various $SU(3)$ gauge groups.

We conclude that axigluon models (or their coloron \cite{Chivukula:1996yr} and topgluon \cite{Hill:1991at} cousins) are unlikely to be the source of the observed top quark asymmetry. Finally, we note that {\it any} theory of
massive colored bosons invoked to explain the anomalous top quark forward-backward asymmetry must 
satisfy the
flavor constraints we have discussed in this note.

%%%%%%%%%%%%%%%%%%%

\section{Acknowledgements}

This work was supported in part by the US National Science Foundation under grants PHY-0354226, 
PHY-0854889, and PHY-0855561.  RSC and EHS gratefully acknowledge the support of the Aspen Center for Physics where part of this work was completed.  CPY thanks K.~Wang for helpful discussions.

\end{document}